\documentstyle [12pt] {article}
\begin{document}
\pagestyle{empty}
\rightline{\vbox{
	\halign{&#\hfil\cr
	&NUHEP-TH-93-1\cr
	&January 1993\cr}}}
\bigskip
\bigskip
\bigskip
{\Large\bf
\centerline{Neutrino Energy Loss from the Plasma Process}
\centerline{at all Temperatures and Densities}}
\bigskip
\normalsize
\centerline{Eric Braaten and Daniel Segel}
\centerline{\sl Department of Physics and Astronomy, Northwestern University,
    Evanston, IL 60208}
\bigskip
\begin{abstract}
We present a unified approach which is accurate at all
temperatures and densities for calculating
the energy loss from a stellar plasma due to the plasma process,
the decay of photons and plasmons into neutrino pairs.
To allow efficient numerical calculations,
an analytic approximation to the dispersion
equations for photons and plasmons is developed.
It is correct to order $\alpha$ in the
classical, degenerate, and relativistic limits for all momenta $k$
and is correct at small $k$ for all temperatures and electron densities.
Within the same approximations, concise expressions
are derived for the transverse, longitudinal,
and axial vector components of the neutrino emissivity.
\end{abstract}
\vfill\eject\pagestyle{plain}\setcounter{page}{1}

The emission of neutrinos can be an important energy loss mechanism for
very hot or dense stars.
The collective effects of the stellar
plasma can significantly alter the production rate of neutrinos.
The most dramatic example is the ``plasma process'',
the decay of photons and plasmons into neutrino pairs,
a process that owes its very existence to plasma effects.
It was pointed out by Adams, Ruderman, and Woo \cite{arw}
in 1963 that the plasma process could be the
dominant energy loss mechanism for very hot and dense stars.
It was recently shown that the formulae for the plasma
process that have been used in all previous work are inaccurate at
relativistic temperatures and electron densities \cite{eba},
underestimating the
emissivity by a factor as large as 3.185.
The purpose of this paper is to provide a unified treatment
of the plasma process that is accurate at all temperatures and densities
and also allows efficient numerical calculations.
We introduce a simple analytic
approximation to the dispersion equations for  photons and plasmons
which becomes exact in the
classical limit, the degenerate limit, and the relativistic limit,
and interpolates smoothly between these limits.  Within the same
approximation, we obtain simple expressions for the transverse, longitudinal,
and axial vector components of the neutrino emissivity from the
plasma process.  The derivation of the analytic
dispersion equations for photons and plasmons  is presented
in Appendix A.  The effective neutrino-photon interaction
that is responsible for the plasma process is discussed in Appendix B and
the decay rate of a photon or plasmon into a neutrino pair is calculated
in Appendix C.

Because a plasma contains mobile charged particles,
an electromagnetic wave propagating through the plasma consists
of coherent vibrations of both the electromagnetic field and the density
of charged particles.
These coherent vibrations behave qualitatively differently
{}from electromagnetic waves in the vacuum in that there are
longitudinal waves as well as transverse
waves, and they propagate at less than the speed of light.
The quantization of the electromagnetic waves in  a plasma
gives rise to a spin-1 particle with 1 longitudinal and 2 transverse
spin polarizations.  It is common in the literature to refer to all 3
polarization states as ``plasmons'',
to emphasize that their dispersion relations
depend on the properties of the plasma.  The longitudinal and transverse
modes are then awkwardly labelled ``longitudinal plasmons''
and ``transverse plasmons''.  While the longitudinal mode owes its
very existence to the plasma, the transverse mode simply
has its dispersion relation at low frequencies modified by the plasma.
For this reason and for the sake of concise terminology,
the label ``plasmon'' will be reserved in this paper for the longitudinal
mode, and the transverse modes will be called ``photons''
whether they are propagating in a plasma or in the vacuum.

The dispersion relations for photons and
for plasmons depend on the temperature $T$ and
the net density $n_e$ of electrons minus positrons.
The general expression for the net electron density
as a function of $T$ and the electron chemical potential $\mu$ is
\begin{equation} {
n_e(T,\mu) \;=\; {1 \over \pi^2} \int_0^\infty dp \; p^2
\Bigg( n_F(E) \;-\; {\bar n}_F(E) \Bigg) \;.
\;\;EB} \label{ne}	\end{equation}
Throughout this paper, we use units in which $\hbar = c = k_B = 1$.
The integral in (\ref{ne}) is over the momentum $p$ of electrons or positrons,
$E=\sqrt{p^2 + m_e^2}$ is their energy,
and $m_e$ is the electron mass.  The Fermi distributions for
electrons and positrons respectively are
\begin{equation} {
n_F(E) \;=\; {1 \over e^{(E-\mu)/T} + 1} \;,
} \label{nF} \end{equation}
\begin{equation} {
{\bar n}_F(E) \;=\; {1 \over e^{(E+\mu)/T} + 1} \;.
} \label{nFbar} \end{equation}
Given the net electron density $n_e$,
the chemical potential is determined by inverting (\ref{ne})
to obtain $\mu(T,n_e)$ as a function of $T$ and $n_e$.

In the vacuum, photons cannot decay into neutrino pairs because
they are massless.  Their dispersion relation
is $\omega^2 = k^2$, and the phase space available for the decay
is proportional to $\omega^2 - k^2 = 0$.  The qualitative behavior of
the dispersion relations in a plasma is shown in Figure 1.
The upper and lower solid curves are the dispersion relations
$\omega_t(k)$ for photons and $\omega_\ell(k)$ for plasmons at a temperature
of $T = 10^{11}$ K = 8.6 MeV and an electron density corresponding to
$\rho/\mu_e = 10^{12} \; {\rm g/cm}^3$.  (The quantity
$\rho/\mu_e$ is the mass density of protons in the plasma and
is related to the net electron density $n_e$ by
$\rho/\mu_e = m_p n_e$, where $m_p$ is the proton mass.)
As $k \rightarrow 0$, the dispersion relations $\omega_t(k)$ and
$\omega_\ell(k)$
both approach the plasma frequency $\omega_p$, which is given by
\begin{equation} {
\omega_p^2 \;=\; {4 \alpha \over \pi} \int_0^\infty dp \; {p^2 \over E}
\left( 1 - {1 \over 3} v^2 \right)
\Bigg( n_F(E) \;+\; {\bar n}_F(E) \Bigg) \;,
} \label{wpl} \end{equation}
where $v = p/E$ is the velocity of the electrons or positrons.
At large $k$, the behavior of the photon dispersion relation is
$\omega_t(k)^2 \rightarrow k^2 + m_t^2$, where $m_t$ is the
transverse photon mass:
\begin{equation} {
m_t^2 \;=\; {4 \alpha \over \pi} \int_0^\infty dp \; {p^2 \over E}
\Bigg( n_F(E) \;+\; {\bar n}_F(E) \Bigg) \;.
} \label{mt} \end{equation}
Comparison with (\ref{wpl}) reveals that the transverse mass
lies in the range $\omega_p \le m_t \le \sqrt{3/2} \; \omega_p$.
As $k$ increases, the dispersion relation $\omega_\ell(k)$ for plasmons
eventually crosses the light cone $\omega = k$ at a point $k_{max}$
given by
\begin{equation} {
k_{max}^2 \;=\; {4 \alpha \over \pi} \int_0^\infty dp \; {p^2 \over E}
\left( {1 \over v} \log {1 + v \over 1 - v} \;-\; 1 \right)
\Bigg( n_F(E) \;+\; {\bar n}_F(E) \Bigg) \;.
} \label{kmax} \end{equation}
It satisfies $\omega_p \le k_{max} < \infty$ and
represents the maximum momentum for which a plasmon can propagate.
As discussed in Appendix A.2, the expressions (\ref{wpl}), (\ref{mt}),
and (\ref{kmax}) for $\omega_p$, $m_t$, and $k_{max}$ are correct to first
order
in the electromagnetic fine structure constant $\alpha \simeq 1/137.036$.
Since the photon dispersion relation satisfies
$\omega_t(k) > k$ for all $k$ and the plasmon dispersion relation
satisfies $\omega_\ell(k) > k$ for $k < k_{max}$, both photons and plasmons
can decay into neutrino pairs.

The production rate of neutrino pairs from the plasma process is sensitive
to the precise form of the photon and plasmon dispersion relations.
In 1961, Tsytovich \cite{tsy}
wrote down general integral equations for the dispersion relations
which include the effects of electrons in the plasma to first
order in $\alpha$. In their pioneering work on the plasma process in 1963,
Adams, Ruderman, and Woo \cite{arw} repeated the general
integral equations, but
they used the following simple dispersion relations in their numerical work:
\begin{equation} {
\omega_t(k)^2 \;=\; \omega_p^2 \;+\; k^2 \;,
\quad 0 \leq k < \infty \;,
} \label{wtarw} \end{equation}
\begin{equation} {
\omega_\ell(k)^2 \;=\; \omega_p^2 \;,
\quad 0 \leq k < \omega_p \;.
} \label{wlarw} \end{equation}
At $k = 0$, these dispersion
relations have the correct value $\omega_p$ at all
temperatures and electron densities.  At nonzero $k$, they
are accurate only at temperatures and densities where the
electrons are nonrelativistic.  In 1967,
Baudet, Petrosian, and Salpeter \cite{bps} improved on the dispersion
relations (\ref{wtarw}) and (\ref{wlarw}) by including
the first relativistic correction:
\begin{equation} {
\omega_t^2 \;=\; \omega_p^2 \;+\; k^2 \;+\; {\omega_1^2 \over 5} \; {k^2 \over
\omega_t^2} \;,
\quad 0 \leq k < \infty \;,
} \label{wtbps} \end{equation}
\begin{equation} {
\omega_\ell^2 \;=\; \omega_p^2 \;+\; {3 \omega_1^2 \over 5} \; {k^2 \over
\omega_\ell^2} \;,
\quad 0 \leq k < \sqrt{\omega_p^2 + 3 \omega_1^2 / 5} \;,
} \label{wlbps} \end{equation}
where $\omega_1$ is given by the integral
\begin{equation} {
\omega_1^2 \;=\; {4 \alpha \over \pi} \int_0^\infty dp \; {p^2 \over E}
\left( { 5 \over 3} v^2 - v^4 \right)
\Bigg( n_F(E) \;+\; {\bar n}_F(E) \Bigg) \;.
} \label{wone} \end{equation}
It lies in the range $0 \le \omega_1 \le \omega_p$.
At small $k$, the dispersion relations (\ref{wtbps}) and (\ref{wlbps})
have the correct behavior to order $k^2$ at all temperatures and densities, but
they are not accurate at large $k$ if electrons are relativistic.

In the subsequent 24 years, the dispersion relations (\ref{wtarw}) for photons
and (\ref{wlbps}) for plasmons were used in all calculations of the plasma
process \cite{dad,mki,sswmt}.
It was pointed out by Braaten \cite{eba} in 1991 that they
could lead to significant errors at temperatures or densities
where electrons become relativistic.  In the relativistic limit,
the correct dispersion relations are the solutions to the
following transcendental equations:
\begin{equation} {
\omega_t^2 \;=\; k^2 \;+\; \omega_p^2 \; {3 \omega_t^2 \over 2 k^2}
\left( 1 \;-\; {\omega_t^2 - k^2 \over \omega_t^2}
	{\omega_t \over 2 k} \log{\omega_t + k \over \omega_t - k} \right) \;,
\quad 0 \leq k < \infty \;,
} \label{wtrel} \end{equation}
\begin{equation} {
\omega_\ell^2 \;=\; \omega_p^2 \; {3 \omega_\ell^2 \over k^2}
\left({\omega_\ell \over 2 k} \log{\omega_\ell+k \over \omega_\ell-k} \;-\; 1
\right) \;,
\quad 0 \leq k < \infty \;.
} \label{wlrel} \end{equation}
These dispersion relations were first derived by Silin
in 1960 using kinetic theory \cite{sil}, and were rederived in 1982 by Klimov
and by Weldon using thermal field theory methods \cite{klw}.
Expanding the right sides of (\ref{wtrel}) and (\ref{wlrel})
in powers of $k$ and using
the fact that $\omega_1 = \omega_p$ in the relativistic limit, one finds that
they agree with (\ref{wtbps}) and (\ref{wlbps}) to order $k^2$.
They differ significantly at large $k$.
For example, the transverse mass in the relativistic limit
is $m_t = \sqrt{3/2} \; \omega_p$, while (\ref{wtbps}) gives the value
$\sqrt{6/5} \; \omega_p$.  Also, the maximum plasmon momentum approaches
infinity in the relativistic limit, while (\ref{wlbps}) gives a
value of $\sqrt{8/5} \; \omega_p$ for $k_{max}$.

In a subsequent paper by Itoh et al. \cite{imhk}, the energy loss from the
plasma process was calculated for a relativistic plasma
using the dispersion relations for a degenerate plasma at zero temperature.
The use of the zero-temperature dispersion relations is valid
only if $T$ is negligible compared to the electron Fermi energy.
Since this condition is not satisfied by the hottest stellar plasmas, the
calculation of Ref. \cite{imhk} can be improved by properly taking into account
the effects of temperature on the dispersion relations.
The general equations for the dispersion relations
at zero temperature were obtained in closed form by
Jancovici in 1962 \cite{janc} and used in the calculations of
Ref. \cite{imhk}, but they are too lengthy to reproduce here.
Jancovici also gave simplified equations for the dispersion relations:
\begin{equation} {
\omega_t^2 \;=\; k^2 \;+\; \omega_p^2 \; {3\omega_t^2 \over 2 v_F^2 k^2}
\left( 1 - {\omega_t^2 - v_F^2 k^2 \over \omega_t^2}
{\omega_t \over 2 v_F k} \log {\omega_t + v_F k \over \omega_t - v_F k}
\right) \;,
\quad 0 \leq k < \infty \;,
} \label{wtdeg} \end{equation}
\begin{equation} {
\omega_\ell^2 \;=\; \omega_p^2 \; {3 \omega_\ell^2 \over v_F^2 k^2}
\left( {\omega_\ell \over 2 v_F k} \log
{\omega_\ell + v_F k \over \omega_\ell - v_F k} - 1 \right) \;,
\quad 0 \leq k < k_{max} \;,
} \label{wldeg} \end{equation}
where $v_F = p_F/E_F$ is the Fermi velocity, $E_F$ is the Fermi energy,
and $p_F = \sqrt{E_F^2 - m_e^2}$ is the Fermi momentum.
The maximum plasmon momentum is
\begin{equation} {
k_{max} \;=\; \left[ {3 \over v_F^2}
\left( {1 \over 2 v_F} \log {1 + v_F \over 1 - v_F}  - 1 \right)
\right]^{1/2} \omega_p \;.
} \label{kmdeg} \end{equation}
The dispersion equations (\ref{wtdeg}) and (\ref{wldeg}) have been
rederived using thermal field theory methods \cite{alt}.
In Ref. \cite{janc}, these simple dispersion equations
were only claimed to be valid for $k << p_F$ and $\omega << E_F$.
However, as shown in Appendix A.2, they are in fact valid for all $k$.
They take into account
correctly all effects of electrons in the plasma to first order in
$\alpha$.  Thus they could have been used in the calculations of
Ref. \cite{imhk} without any loss of accuracy.
Expanding the right side of (\ref{wtdeg}) and (\ref{wldeg}) for
small $k$ and using the fact that $\omega_1 = v_F \omega_p$ in the degenerate
limit,
we see that they agree with (\ref{wtbps})
and (\ref{wlbps}) to order $k^2$.  In the relativistic limit where
$v_F \rightarrow 1$, (\ref{wtdeg}) and (\ref{wldeg}) reduce to the
relativistic dispersion relations (\ref{wtrel}) and (\ref{wlrel}).

Aside from the numerical complications of solving integral equations,
the general dispersion relations for photons and plasmons
given in Ref. \cite{tsy} also
suffer from a technical difficulty in that
$\omega_t(k)$ and $\omega_\ell(k)$ become complex-valued if the temperature
or the Fermi energy is sufficiently high.  The imaginary parts appear when
$\omega_p$ exceeds the threshold $2m_e$ for decay
of a photon into an electron-positron pair in the vacuum.
This threshold is unphysical, because the rest mass of an electron
in a relativistic plasma is significantly greater than in the vacuum.
The effects of the plasma on the electron and photon
dispersion relations are such that the decay of a photon
or plasmon into an $e^+ e^-$ pair is always forbidden by
energy and momentum conservation \cite{ebb}.
The problem of complex-valued dispersion relations persists
in the degenerate limit, although the
threshold is increased to $\omega > E_F + m_e$, corresponding to the
production of a positron at rest and an electron at the Fermi surface.
The effects of the unphysical process $\gamma \rightarrow e^+ e^-$
may be numerically small,
but it would be preferable to have dispersion relations
{}from which they are absent altogether.

In Appendix A.2, we derive dispersion relations for photons and plasmons
which receive no contributions from the unphysical
process $\gamma \rightarrow e^+ e^-$.  This
is achieved without any sacrifice in accuracy:  all effects from
electrons and positrons in the plasma are still included
to first order in $\alpha$. The resulting expressions (\ref{wpl}),
(\ref{mt}), and (\ref{kmax}) for the plasma frequency, the transverse
photon mass, and the maximum plasmon momentum are much simpler than,
but just as accurate as, the corresponding expressions that follow directly
{}from the general integral equations.
The dispersion relations remain real-valued at all temperatures and
electron densities.
Their behavior at small $k$ is consistent with (\ref{wtbps}) and
(\ref{wlbps}), they reduce to (\ref{wtrel}) and (\ref{wlrel})
in the relativistic limit, and they reduce to (\ref{wtdeg}) and (\ref{wldeg})
in the degenerate limit.

The dispersion equations given in Appendix A.2
are integral equations involving
1-dimensional integrals over the momenta of electrons and positrons.
Only in the classical limit, the degenerate limit, and the relativistic
limit can the integrals be evaluated analytically.
At intermediate temperatures and electron densities, calculating the
dispersion relation $\omega_t(k)$ or $\omega_\ell(k)$ requires finding,
for each value of
$k$, the zero in $\omega$ of an $\omega$-dependent integral.
The numerical solution of such a complicated dispersion equation is
too inefficient for many applications, such as calculating the energy loss
{}from the plasma process.
We have therefore developed an approximation to these dispersion
relations that is remarkably accurate at all temperatures and electron
densities, but is simple enough to be used for practical calculations.
The derivation of these dispersion
relations is presented in Appendix A.4.  They are correct to order $\alpha$ in
the classical limit, the degenerate limit, and the relativistic  limit,
and they provide a smooth interpolation to intermediate temperatures and
electron densities.  They
are also correct to order $k^2$ at small $k$ for all temperatures
and electron densities.

Our approximate dispersion relations are as simple as the degenerate
dispersion relations given in (\ref{wtdeg}) and (\ref{wldeg}).
Given the plasma frequency (\ref{wpl}) and
the freqency $\omega_1$ given by (\ref{wone}), we define a parameter $v_*$:
\begin{equation} {
v_* \;=\; {\omega_1 \over \omega_p} \;.
} \label{vstar} \end{equation}
It lies in the range $0 \le v_* \le 1$, and has
the intuitive interpretation of
a typical velocity of electrons in the plasma.
Our approximate dispersion relations $\omega_t(k)$ and $\omega_\ell(k)$
are obtained by solving the following equations which depend on $v_*$:
\begin{equation} {
\omega_t^2 \;=\; k^2 \;+\; \omega_p^2 \; {3\omega_t^2 \over 2 v_*^2 k^2}
\left( 1 - {\omega_t^2 - v_*^2 k^2 \over \omega_t^2}
{\omega_t \over 2 v_* k} \log {\omega_t + v_* k \over \omega_t - v_* k}
\right) \;,
\quad 0 \leq k < \infty \;,
} \label{wtstar} \end{equation}
\begin{equation} {
\omega_\ell^2 \;=\; \omega_p^2 \; {3 \omega_\ell^2 \over v_*^2 k^2}
\left( {\omega_\ell \over 2 v_* k} \log
{\omega_\ell + v_* k \over \omega_\ell - v_* k} - 1 \right) \;,
\quad 0 \leq k < k_{max} \;.
} \label{wlstar} \end{equation}
The maximum plasmon momentum is
\begin{equation} {
k_{max} \;=\; \left[ {3 \over v_*^2}
\left( {1 \over 2 v_*} \log {1 + v_* \over 1 - v_*}  - 1 \right) \right]^{1/2}
\omega_p \;.
} \label{kmstar} \end{equation}
The transverse photon mass is
\begin{equation} {
m_t \;=\; \left[ {3 \over 2 v_*^2}
\left( 1 - {1 - v_*^2 \over 2 v_*} \log {1 + v_* \over 1 - v_*} \right)
\right]^{1/2} \omega_p \;.
} \label{mtstar} \end{equation}
These expressions satisfy $2 m_t^2 + (1 - v_*^2) k_{max}^2 = 3 \omega_p^2$.
The dispersion equations (\ref{wtstar}) and (\ref{wlstar})
are correct to order $\alpha$ for all $k$
in three limiting cases  that are studied in Appendix A.3.
In the classical limit, (\ref{vstar}) gives $v_* = \sqrt{5T/m_e}$.
The dispersion equations (\ref{wtstar}) and (\ref{wlstar}), when expanded
to first order in $v_*^2$, agree with (\ref{wtbps}) and (\ref{wlbps}).
In the relativistic limit, $v_* = 1$ and
(\ref{wtstar}) and (\ref{wlstar}) reduce to
the relativistic dispersion equations (\ref{wtrel}) and (\ref{wlrel}).
In the degenerate limit, $v_*$ is equal to the Fermi velocity $v_F$ and
the dispersion equations reduce to (\ref{wtdeg}) and (\ref{wldeg}).
Our approximate dispersion relations
are also correct to order $k^2$ at all temperatures and electron densities.
Expanding the right side of (\ref{wlstar}) and (\ref{wtstar}) in powers
of $k$, we find that to order $k^2$ they reduce to (\ref{wtbps})
and (\ref{wlbps}).
At large $k$, the dispersion equations (\ref{wtstar}) and (\ref{wlstar})
are correct to order $\alpha$
only in the classical, degenerate, and relativistic  limits.  For example,
the expressions (\ref{kmstar}) and (\ref{mtstar}) for $k_{max}$ and $m_t$
are not identically equal to the general expressions (\ref{kmax}) and
(\ref{mt}), but the differences are found empirically to be
remarkably small.

To calculate the energy loss from the plasma process,
we require the photon and plasmon dispersion relations $\omega_t(k)$ and
$\omega_\ell(k)$,
the corresponding residue factors $Z_t(k)$ and $Z_\ell(k)$,
and the axial polarization function $\Pi_A(\omega,k)$
evaluated at the photon dispersion relation $\omega = \omega_t(k)$.
General expressions for the residue factors
and for $\Pi_A(\omega,k)$ are given in Appendices A and B.
The use of these general expressions in calculations such
as the energy loss is very cumbersome.
Fortunately, the methods used to derive the approximate dispersion
relations (\ref{wtstar}) and (\ref{wlstar}) can also be used to derive
compact analytic expressions for $Z_t$, $Z_l$, and $\Pi(\omega_t,k)$
that are correct to order $\alpha$
in the classical limit, the degenerate limit, and the relativistic limit
and are also correct for small $k$ at all temperatures and electron densities.
As shown in Apppendix A.4,
the transverse and longitudinal residue factors can  be approximated by
\begin{equation} {
Z_t(k) \;=\;
{ 2 \omega_t^2 (\omega_t^2 - v_*^2 k^2) \over  3 \omega_p^2 \omega_t^2
+ (\omega_t^2 + k^2) (\omega_t^2 - v_*^2 k^2) - 2 \omega_t^2
(\omega_t^2 - k^2) } \;,
} \label{Ztstar} \end{equation}
\begin{equation} {
Z_\ell(k) \;=\;
{ 2 (\omega_\ell^2 - v_*^2 k^2) \over 3 \omega_p^2 -
(\omega_\ell^2 - v_*^2 k^2) } \;.
} \label{Zlstar} \end{equation}
As shown in Appendix B,
the axial polarization function $\Pi(w,k)$ evaluated at
$\omega = \omega_t(k)$ can be approximated by
\begin{equation} {
\Pi_A(\omega_t,k) \;=\; \omega_A \; k \;
{\omega_t^2 - k^2 \over \omega_t^2 - v_*^2 k^2} \;
{3 \omega_p^2 - 2 (\omega_t^2 - k^2) \over \omega_p^2} \;.
} \label{PiAstar} \end{equation}
Its behavior as $k \rightarrow 0$ is
$\Pi_A(\omega_t,k) \rightarrow \omega_A k$,
where the coefficient $\omega_A$ is
\begin{equation} {
\omega_A \;=\; {2 \alpha \over \pi} \int_0^\infty dp \; {p^2 \over E^2}
\left( 1 - {2 \over 3} v^2 \right)
\Bigg( n_F(E) \;-\; {\bar n}_F(E) \Bigg) \;.
} \label{wA} \end{equation}

The emissivity $Q$ of a plasma is the rate of energy loss per unit volume.
To calculate the emissivity from the plasma process, one must first calculate
the rates $\Gamma_t(k)$  and $\Gamma_\ell(k)$ for the decay of
a photon and a plasmon of momentum $k$ into a $\nu {\bar \nu}$ pair.
These rates are calculated in Appendix C under the assumption that the
number density of neutrinos and antineutrinos remains negligible.
The emissivity is then obtained by integrating
over the phase space of the photon or plasmon, weighted by the number density
and by the energy.
Summing over the 3 polarization states of photons and plasmons
and over the neutrino types $\nu = \nu_e, \nu_\mu, \nu_\tau$,
the total emissivity is
\begin{equation} {
Q \;=\; \sum_\nu \int {d^3k \over (2 \pi)^3}
\; \Bigg( 2 \; n_B(\omega_t(k)) \; \omega_t(k) \; \Gamma_t(k)
\;+\; n_B(\omega_\ell(k)) \; \omega_\ell(k) \; \Gamma_l(k) \Bigg) \;.
} \label{Qdef} \end{equation}
The number densities of photons and plasmons
are given by the Bose distribution:
\begin{equation} {
n_B(\omega) \;=\; {1 \over e^{\omega/T} - 1} \;.
} \label{bose} \end{equation}
The total emissivity $Q$ can be separated
into the vector ($Q_T$) and axial vector ($Q_A$) components
of the photon contribution and the plasmon contribution ($Q_L$):
\begin{equation} {
Q_T \;=\; 2 \left( \sum_\nu C_V^2 \right) {G_F^2 \over 96 \pi^4 \alpha}
\int_0^\infty dk \; k^2 \; Z_t(k)
	\left( \omega_t(k)^2 - k^2 \right)^3 n_B(\omega_t(k)) \;,
} \label{QT} \end{equation}
\begin{equation} {
Q_A \;=\; 2  \left( \sum_\nu C_A^2 \right) {G_F^2 \over 96 \pi^4 \alpha}
\int_0^\infty dk \; k^2 \; Z_t(k)
\left( \omega_t(k)^2 - k^2 \right) \;
\Pi_A(\omega_t(k),k)^2 \; n_B(\omega_t(k)) \;,
} \label{QA} \end{equation}
\begin{equation} {
Q_L \;=\;  \left( \sum_\nu C_V^2 \right)
{G_F^2 \over 96 \pi^4 \alpha}
\int_0^{k_{max}} dk \; k^2 \; Z_\ell(k) \; \omega_\ell(k)^2
\left( \omega_\ell(k)^2 - k^2 \right)^2 n_B(\omega_\ell(k)) \;.
} \label{QL} \end{equation}
The coefficients $C_V$ and $C_A$ depend on the neutrino type and
are given in Appendix B.
The combinations that arise in the plasma process are
\begin{equation} {
\sum_\nu C_V^2 \;=\;
{3 \over 4} \;-\; 2 \sin^2 \theta_W \;+\; 12 \sin^4 \theta_W
	\; \approx \;  0.911 \;,
} \label{CVsq} \end{equation}
\begin{equation} {
\sum_\nu C_A^2 \;=\; {3 \over 4} \;.
} \label{CAsq} \end{equation}
The formulas for $Q_T$ and $Q_A$ were first correctly given
up to the factor of $\sum C_V^2$ in Ref. \cite{tz}. The correct values
of $C_V$ were first included in Ref. \cite{dad}.
The formula (\ref{QL}) for $Q_L$ differs from that given in
Refs. \cite{eba} and \cite{imhk} because the definition of the
longitudinal residue factor $Z_\ell(k)$
differs by a factor of $(\omega_\ell^2-k^2)/\omega_\ell^2$.
The formula (\ref{QA}) for $Q_A$ supercedes that given in Ref. \cite{kim},
in which  the factor $Z_t(k)$ was omitted and the photon was assumed to
satisfy the simple dispersion relation (\ref{wtarw}).

The important momentum scales in the emissivity integrals appearing in
(\ref{QT}), (\ref{QA}), and (\ref{QL})
are the plasma frequency $\omega_p$ (which enters into the dispersion
relations, the residue factors, and $\Pi_A(\omega_t,k)$),
and the temperature $T$ (which occurs in the Bose distribution).
The integrals can be simplified in the limiting cases
$T >> \omega_p$ and $\omega_p << T$.
We first consider the high temperature limit $T >> \omega_p$.
The integral (\ref{QT}) for the transverse
emissivity is dominated by momenta $k$ of order $T$.
Since $T >> \omega_p$, the photon dispersion relation can be approximated
by $\omega_t^2 = k^2 + m_t^2$.  We can therefore replace the factors of
$\omega_t(k)^2 - k^2$ in (\ref{QT}) by $m_t^2$, and set
$\omega_t = k$ everywhere else in the integrand.
The integral over $k$ can then be evaluated analytically in
terms of the Riemann zeta function: $\zeta_3 \simeq 1.202057$.
The result is
\begin{equation} {
Q_T \;\rightarrow\;
\left( \sum_\nu C_V^2 \right) {G_F^2 \over 96 \pi^4 \alpha}
\; 4 \zeta_3 \; m_t^6 \; T^3 \;.
} \label{QThot} \end{equation}
The integral (\ref{QA}) for the axial vector emissivity is also dominated by
momenta $k$ of order $T$.
Applying similar approximations as were used for the transverse emissivity,
it can also with some effort be evaluated analytically:
\begin{equation} {
Q_A \;\rightarrow\;
\left( \sum_\nu C_A^2 \right) {G_F^2 \over 96 \pi^4 \alpha}
\; 2 \; \left[
{3 \over v_*^2} \left({1 \over 2v_*}\log{1+v_* \over 1-v_*} - 1 \right)
\right]^2
\; m_t^6 \; \omega_A^2 \; T \; \log {2T \over m_t} \;.
} \label{QAhot} \end{equation}
Up to logarithmic factors,
the axial vector emissivity is suppressed relative to $Q_T$ by a factor
of $\omega_A^2/T^2$.  In the limit $T >> \omega_p$,
the integral (\ref{QL}) for the longitudinal emissivity
is dominated by momenta $k$ of order $\omega_p$.
This is obvious in the nonrelativistic limit, because the upper limit $k_{max}$
is of order $\omega_p$.  In the relativistic limit
where $k_{max}$ approaches infinity, the factors $Z_\ell(k)$
and $(\omega_\ell(k)^2 - k^2)^2$ both approach zero very rapidly for
$ k >> \omega_p$,
providing a cutoff on the integral of order $\omega_p$.
If $k$ is of order $\omega_p$, then $\omega_\ell(k)$ is also of order
$\omega_p$.
Using $\omega_\ell(k) << T$, the Bose factor in (\ref{QL}) can be simplified
to $n_B(\omega_\ell) \rightarrow T/\omega_\ell$.  The integral
can still not be calculated analytically, but by dimensional analysis
it must be proportional to $\omega_p^8 \; T$.  The longitudinal emissivity
then has the form
\begin{equation}{
Q_L \;\rightarrow\;  \left( \sum_\nu C_V^2 \right)
{G_F^2 \over 96 \pi^4 \alpha} \; A(v_*) \; \omega_p^8 \; T \;.
} \label{QLhot} \end{equation}
The coefficient $A(v_*)$ varies slowly from $8/105$ in the
nonrelativistic limit ($v_* \rightarrow 0$) to 0.349 in the
relativistic limit ($v_* \rightarrow 1$) \cite{eba}.
The longitudinal emissivity is suppressed relative to the
transverse emissivity by a factor of $\omega_p^2/T^2$.

We next consider the low temperature limit $T << \omega_p$.
The integrals in (\ref{QT}) and (\ref{QA}) are dominated by momenta
$k << \omega_p$.  The Bose distribution
can therefore be approximated by a Gaussian in $k$ :
$n_B(\omega_t(k)) \rightarrow e^{-\omega_p/T} exp(-\omega_t''(0) k^2/2T)$,
where $\omega_t''(0) = (1 + v_*^2/5)/\omega_p$.
Everywhere else in the integrands,
we can set $\omega_t = \omega_p$ and ignore $k$ relative to $\omega_p$.
The integrals can then be evaluated analytically,
with the results
\begin{equation} {
Q_T \;\rightarrow\;
\left( \sum_\nu C_V^2 \right) {G_F^2 \over 96 \pi^4 \alpha} \;
\sqrt{2 \pi} \; \left( 1 + {1 \over 5} v_*^2 \right)^{-3/2}
\; \omega_p^{15/2} \; T^{3/2} \; e^{-\omega_p/T} \;,
} \label{QTcold} \end{equation}
\begin{equation} {
Q_A \;\rightarrow\;
\left( \sum_\nu C_A^2 \right) {G_F^2 \over 96 \pi^4 \alpha} \;
3 \sqrt{2 \pi} \; \left( 1 + {1 \over 5} v_*^2 \right)^{-5/2}
\; \omega_p^{9/2} \; \omega_A^2 \; T^{5/2} \; e^{-\omega_p/T} \;.
} \label{QAcold} \end{equation}
The axial vector emissivity is suppressed relative to
$Q_T$ by a factor of $\omega_A^2 T/\omega_p^3$.
A similar approximation can be applied to the longitudinal emissivity
provided that $T << v_*^2 \omega_p$.  The Bose distribution in (\ref{QL})
can then be approximated by a Gaussian in $k$:
$n_B(\omega_\ell(k)) \rightarrow e^{-\omega_p/T} exp(-\omega_\ell''(0)
k^2/2T)$,
where $\omega_\ell''(0) = (3/5)v_*^2/\omega_p$.  The expression for the
emissivity
then reduces to
\begin{equation} {
Q_L \;\rightarrow\;  \left( \sum_\nu C_V^2 \right)
{G_F^2 \over 96 \pi^4 \alpha} \;
\sqrt{{\pi \over 2}} \; \left( {3 \over 5} v_*^2 \right)^{-3/2}
\; \omega_p^{15/2} \; T^{3/2} \; e^{-\omega_p/T} \;.
} \label{QLcold} \end{equation}
At relativistic electron densities ($v_* \rightarrow 1$),
the longitudinal emissivity
is smaller than the transverse emissivity only by a factor of $\sqrt{2}$.

In Figure 2, the components $Q_T$, $Q_A$, and $Q_L$ of the emissivity
are shown as a function of the proton mass density $\rho/\mu_e = m_p n_e$
at a temperature $T = 10^{11}$ K = 8.6 MeV.
Also shown as dotted lines are the
corresponding emissivities calculate using the 0-temperature dispersion
relations as in Ref. \cite{imhk}.  While they give the same results at
high densities, there are significant discrepancies at the lowest
densities considered in Figure 2.  The shapes of the dispersion relations
at $T=0$ and $T = 10^{11} K$ are very similar, since the electrons are
relativistic over the entire range of densities shown.  The discrepancies in
Figure 2 therefore arise primarily from the difference
in the value of the plasma frequency.
At the highest densities shown, $\omega_p$ is determined primarily
by the Fermi energy, so the 0-temperature dispersion relations
provide a good approximation.
At the lowest densities shown, $\omega_p$ is determined primarily
by the temperature.  The emissivities $Q_T$ and $Q_L$ therefore
become almost independent of the density, as can be seen in Figure 2.

The formulas for the dispersion relations and the neutrino emissivity
that have been presented above were derived for
a plasma of electrons and  positrons in a uniform positively
charged background that cancels the net charge of the electrons and positrons.
In a stellar plasma, the cancelling charge is actually provided by
protons and heavier ions.  These charged particles will also give
contributions both to the electromagnetic polarization functions and to the
effective photon-neutrino interaction that add to those of the electrons
and positrons.  The effects of protons and heavier ions are negligible for
stellar plasmas as long as they are nonrelativistic.
At the highest densities shown in Figure 2, the protons are relativistic
and their effects are significant.  At $T = 10^{11}$ K and
$\rho/\mu_e = 10^{14} \; {\rm g/cm}^3$,
electrons and protons are degenerate with both having a Fermi momentum
of 238 MeV.  Their contributions to $\omega_p^2$ are proportional to
their Fermi velocities, which is 1 for electrons and 0.246 for protons.
Thus the protons increase the plasma frequency by about $12 \%$.
The resulting effect on the emissivity will be much larger, since it scales
like a large power of the plasma frequency.  The modifications to the
formulas for the emissivities that are required in order to take
into account the effects of protons are given in Appendix C.

We have developed a unified approach for calculating the neutrino
energy loss from the plasma process at all temperatures and densities.
We have introduced compact equations for the dispersion relations
for photons and plasmons that are correct to order $\alpha$ for all $k$
in the classical, degenerate, and relativistic limits
and are also correct to order $k^2$ at small $k$ for all temperatures
and electron densities.  Compact expressions were also obtained for the
other quantities that are required to calculate the transverse, longitudinal,
and axial vector components of the neutrino emissivity.  Most previous
numerical studies of the  energy loss from the plasma process
have concentrated for simplicity on the case where the neutrino
density remains negligible.  In supernova explosions, neutrinos
become trapped inside a neutrinosphere and the neutrino
emissivity can be suppressed by Pauli blocking effects.
Our approach should allow the efficient numerical investigation of
such effects.   It would also be useful to extend our approach to other
neutrino emission processes, to axion emission, and to other
particle physics proceeses that
can play an important role in astrophysics over  enormous ranges
of temperatures and densities.
The investigation of all of these processes would benefit from a unified
treatment that remains accurate
even under extreme variations of temperature and density.

This work was supported in part by the U.S. Department of Energy,
Division of High Energy Physics, under Grant DE-FG02-91-ER40684.
\vfill\eject

\appendix

\section{Photon and plasmon dispersion relations}

\subsection{Dispersion relations to order $\alpha$}

The effects of a plasma on the propagation of photons and plasmons
is determined by the electromagnetic polarization tensor
$\Pi^{\mu \nu}(K)$.
If interactions with electrons and positrons are taken into account to
leading order in the electromagnetic coupling constant $\alpha$,
the polarization tensor is
\begin{eqnarray}
\Pi^{\mu \nu}(K) &=&
 16 \pi \alpha \int {d^3p \over (2 \pi)^3} \; {1 \over 2 E}
\Bigg( n_F(E) \;+\; {\bar n}_F(E) \Bigg)
\nonumber \\
& \times & { P \cdot K (P^\mu K^\nu + K^\mu P^\nu)
	- K^2 P^\mu P^\nu - (P \cdot K)^2 g^{\mu \nu}
\over (P \cdot K)^2 - (K^2)^2/4 } \;,
 \label{Pimunu} \end{eqnarray}
where $K^\mu = (\omega, {\vec k})$, $P^\mu = (E,{\vec p})$,
$K^2 = \omega^2 - k^2$, and $P \cdot K = E \omega - {\vec p} \cdot {\vec k}$.
The integral is over the momentum ${\vec p}$ of electrons and positrons
with energy $E = \sqrt{p^2 + m_e^2}$.
The polarization tensor satisfies $K_\mu \Pi^{\mu \nu}(K) = 0$,
which is a consequence of gauge invariance.  The transverse
and longitudinal polarization functions $Pi_t(\omega,k)$ and
$Pi_\ell(\omega,k)$ are
\begin{equation} {
Pi_t(\omega,k) \;=\;
{1 \over 2} \left( \delta^{ij} - {\hat k}^i {\hat k}^j \right)
\Pi^{ij}(\omega,{\vec k}) \;,
} \label{Pitdef}	\end{equation}
\begin{equation} {
Pi_\ell(\omega,k) \;=\; \Pi^{00}(\omega,{\vec k}) \;.
} \label{Pildef} \end{equation}
They are related to the standard transverse and longitudinal dielectric
functions $\epsilon_t$ and $\epsilon_\ell$ by
$\epsilon_t = 1 - Pi_t/\omega^2$ and
$\epsilon_\ell = 1 - Pi_\ell/k^2$.

In order to construct the effective propagator
$D^{\mu \nu}(\omega,{\vec k})$ for the
electromagnetic field, it is necessary to choose a gauge.  The most
convenient choice for treating plasma effects is the Coulomb gauge
defined by ${\vec \nabla} \cdot {\vec A} = 0$.  In Coulomb gauge, the nonzero
components of the effective propagator are
\begin{equation} {
D^{00}(\omega,{\vec k}) \;=\; {1 \over k^2 - Pi_\ell(\omega,k)} \;,
} \label{Dldef} \end{equation}
\begin{equation} {
D^{ij}(\omega,{\vec k}) \;=\; {1 \over \omega^2 - k^2 - Pi_t(\omega,k)}
\; \left( \delta^{ij} - {\hat k}^i {\hat k}^j \right)\;.
} \label{Dtdef} \end{equation}
The dispersion relations $\omega_t(k)$ for photons and
$\omega_\ell(k)$ for plasmons
are the locations of the poles in the effective propagator:
\begin{equation} {
D^{00}(\omega,{\vec k}) \;\rightarrow\;
{\omega_\ell(k)^2 \over k^2} \; {Z_\ell(k) \over \omega^2 - \omega_\ell(k)^2}
\quad {\rm as} \quad \omega \rightarrow \omega_\ell(k) \;,
} \label{Dlpole} \end{equation}
\begin{equation} {
D^{ij}(\omega,{\vec k}) \;\rightarrow\; {Z_t(k) \over \omega^2 - \omega_t(k)^2}
\; \left( \delta^{ij} - {\hat k}^i {\hat k}^j \right)
\quad {\rm as} \; \omega \rightarrow \omega_t(k) \;.
} \label{Dtpole} \end{equation}
The locations of the poles are independent of the choice of gauge \cite{kkr}.

The equations (\ref{Dlpole}) and (\ref{Dtpole})
also define the residue functions $Z_\ell(k)$ and $Z_t(k)$,
which determine the strength with which a plasmon or photon of momentum $k$
couples to an electromagnetic current:
\begin{equation} {
Z_t(k) \;=\;
\left[ 1 \;-\; {\partial Pi_t \over \partial \omega^2} (\omega_t(k),k)
\right]^{-1} \;,
} \label{Ztdef} \end{equation}
\begin{equation} {
Z_\ell(k) \;=\; {k^2 \over \omega_\ell(k)^2}
\left[ \;-\; {\partial Pi_\ell \over \partial \omega^2} (\omega_\ell(k),k)
\right]^{-1} \;.
} \label{Zldef} \end{equation}
The longitudinal residue $Z_\ell(k)$ defined by (\ref{Zldef})
differs by a factor of $(\omega_\ell^2-k^2)/\omega_\ell^2$ from
that which was used in Refs. \cite{eba} and \cite{imhk}.
The residue of a pole in
$\omega^2$ of $D^{\mu \nu}(\omega,k)$ can be identified as
$\epsilon^\mu({\vec k}) \epsilon^\nu({\vec k})^*$,
where $\epsilon^\mu({\vec k})$ is the polarization
4-vector for the appropriate propagating mode.  These modes are conveniently
labelled by the helicity $\lambda$, which is the component of the angular
momentum in the direction of ${\vec k}$:  $\lambda = 0$ for plasmons,
$\lambda = \pm 1$ for photons.  From (\ref{Dlpole}) and (\ref{Dtpole}),
we identify the polarization 4-vectors to be
\begin{equation} {
\epsilon^\mu({\vec k},\lambda=0) \;=\;
{\omega_\ell(k) \over k}  \sqrt{Z_\ell(k)} \; (1,0)^\mu \;,
} \label{epsl} \end{equation}
\begin{equation} {
\epsilon^\mu({\vec k},\lambda= \pm 1) \;=\;
\sqrt{Z_t(k)} \; \left(0,{\vec \epsilon}_\pm({\vec k})\right)^\mu \;,
} \label{epst} \end{equation}
where ${\vec \epsilon}_+({\vec k})$ and ${\vec \epsilon}_-({\vec k})$ are
orthogonal to ${\vec k}$
and normalized so that ${\vec \epsilon}_\pm({\vec k}) \cdot
{\vec \epsilon}_\pm({\vec k})^* = 1$.
The polarization 4-vectors (\ref{epsl}) and (\ref{epst})
satisfy the Coulomb gauge constraint
\begin{equation} {
{\vec k} \cdot {\vec \epsilon}({\vec k},\lambda) \;=\; 0 \;.
} \end{equation}
They are used in Appendix C to calculate the decay rate of a photon
or plasmon into neutrino pairs.

The dispersion relations $\omega_t(k)$ for photons
and $\omega_\ell(k)$ for plasmons are the solutions to the equations
\begin{equation} {
\omega_t(k)^2 \;=\; k^2 \;+\; Pi_t(\omega_t(k),k) \;,
} \label{wtdef} \end{equation}
\begin{equation} {
\omega_\ell(k)^2 \;=\; {\omega_\ell(k)^2 \over k^2} Pi_\ell(\omega_\ell(k),k)
\;.
} \label{wldef} \end{equation}
If the complete order-$\alpha$ expression
(\ref{Pimunu}) for the polarization tensor is used to calculate
these dispersion relations, they become complex-valued
when the temperature or electron density
is large enough that the plasma frequency $\omega_p$ exceeds $2 m_e$.
The imaginary part is proportional to the decay rate
for a photon or plasmon into an electron-positron pair:
$\gamma \rightarrow e^+ e^-$.
These imaginary parts are unphysical, because the plasma effects that
give the photon a nontrivial dispersion relation also change the
energy-momentum relation for the electron.  These corrections are
such that the decay $\gamma \rightarrow e^+ e^-$ is always
forbidden by energy and momentum conservation \cite{ebb}.
The unphysical effects from this forbidden process also
reveal themselves in the analytic behavior of the real parts of the
dispersion relations.  Thus they can not be eliminated simply
by defining the dispersion relations $\omega_t(k)$ and $\omega_\ell(k)$
to be the real parts of the poles in the effective propagator.

\subsection{Removing effects of electron pair production}

To eliminate
the unphysical effects of the forbidden decay $\gamma \rightarrow e^+ e^-$,
we exploit the fact that there is a separation
of momentum scales between the particles whose propagation is significantly
modified by the plasma and those that are responsible for the
plasma corrections.   For example,
in the relativistic limit, the dominant contributions to
the electromagnetic polarization tensor in (\ref{Pimunu})
come from electrons and positrons with
momentum $p$ of order the temperature $T$ or the Fermi momentum $p_F$,
whichever is larger.  However plasma corrections to the electron
propagator are significant only for momenta that are smaller by
a factor of $\sqrt{\alpha}$.  In the nonrelativistic limit,
plasma corrections to the electron propagator are always negligible.
This justifies the
use of the vacuum energy-momentum relation $E = \sqrt{p^2 + m_e^2}$
in calculating $\Pi^{\mu \nu}$.

The separation of momentum scales also allows us to simplify the expression
(\ref{Pimunu}) for the polarization tensor by dropping the term
$(K^2)^2/4$ in the denominator.
The physical interpretation is that this corresponds to calculating
plasma corrections using the forward scattering amplitudes
for electrons and positrons in the vacuum \cite{gb}.
The mathematical approximation that is required is
\begin{equation} {
|\omega^2 - k^2| \;<<\; 2 E |\omega - {\vec v} \cdot {\vec k}| \;,
} \label{ineq} \end{equation}
where ${\vec v} = {\vec p}/E$ is the velocity of the electron or positron.
We first discuss the case where $k$ is of order $\omega_p$.
Since the dispersion relations $\omega_t$ and $\omega_\ell$ are then also
of order $\omega_p$, the inequality (\ref{ineq}) requires
$\omega_p << E$.  This is always satisfied in the nonrelativistic limit
where $E$ is of order $m_e$.
In the relativistic limit, the inequality $\omega_p << E$
fails to be satisfied only in a region of
the electron or positron phase space whose contribution to the integral
in (\ref{Pimunu}) is down by a factor of $\omega_p^2/T^2$ in the high
temperature limit and a factor of $\omega_p^2/E_F^2$ in the degenerate limit.
In either case, the contribution is suppressed by a factor of $\alpha$.
We next discuss the case in which $k$ is much greater than $\omega_p$.
The solutions to the dispersion relations always have $\omega^2 - k^2$
of order $\omega_p^2$.  Except in the relativistic limit,  the inequality
(\ref{ineq}) then reduces to $\omega_p^2/k << E$, which is even more easily
satisfied than the condition $\omega_p << E$ discussed above.
In the relativistic limit, the inequality (\ref{ineq}) may fail
for electrons with velocity ${\vec v}$ within an angle of order $\omega_p/k$
of the momentum ${\vec k}$, but this region of phase space gives a
contribution to the integral in (\ref{Pimunu}) that is suppressed by at
least a factor of $\alpha$.
Thus the effects of the $(K^2)^2$ term in the
denominator of (\ref{Pimunu}) are always
at most comparable to the order $\alpha^2$ corrections to the polarization
tensor.  It can therefore be dropped without any loss in accuracy.

Dropping the $(K^2)^2$ term in the denominator of
(\ref{Pimunu}) eliminates the effects of the unphysical process
$\gamma \rightarrow e^+ e^-$, so that
the dispersion relations for photons and plasmons
remain real-valued at all temperatures and densities.
It also results in far simpler expressions
for the transverse and longitudinal polarization
functions defined in (\ref{Pitdef}) and (\ref{Pildef}).
The angular integrals can be evaluated analytically, and the
polarization functions reduce to
\begin{equation} {
Pi_t(\omega,k) \;=\; {4 \alpha \over \pi} \int_0^\infty dp \; {p^2 \over E}
\left(  {\omega^2 \over k^2} - {\omega^2 - k^2 \over k^2}
	{\omega \over 2 v k} \log {\omega + v k \over \omega - v k} \right)
\Bigg( n_F(E) \;+\; {\bar n}_F(E) \Bigg) \;,
} \label{Pit}	\end{equation}
\begin{equation} {
Pi_\ell(\omega,k) \;=\; {4 \alpha \over \pi} \int_0^\infty dp \; {p^2 \over E}
\left({\omega \over v k} \log {\omega + v k \over \omega - v k}
	- 1 - {\omega^2 - k^2 \over \omega^2 - v^2 k^2} \right)
\Bigg( n_F(E) \;+\; {\bar n}_F(E) \Bigg) \;.
} \label{Pil} \end{equation}
Inserting (\ref{Pit}) and (\ref{Pil}) into (\ref{wtdef}) and (\ref{wldef})
and solving these equations numerically, one finds that
the dispersion relation $\omega_t(k)$ remains real-valued for
all $k$ and $\omega_\ell(k)$ remains real-valued for those values of $k$
that satisfy $\omega_\ell(k) > k$, corresponding to the timelike
propagation of a plasmon.
As $k \rightarrow 0$, both dispersion relations approach
the plasma frequency, which is given by (\ref{wpl}).
At large $k$, the behavior of the photon dispersion relation
is governed by the transverse photon mass given by (\ref{mt}).
The plasmon dispersion relation crosses the line $\omega = k$
at the point $k_{max}$ given by (\ref{kmax}).

\subsection{Limiting cases}

The polarization functions (\ref{Pit}) and (\ref{Pil})
can be evaluated analytically in three limits: the classical limit,
the degenerate limit, and the relativistic limit.  This allows
the equations (\ref{wtdef}) and (\ref{wldef})
for the dispersion relations to be written in
closed form.  We also obtain analytic expressions for
the plasma frequency (\ref{wpl})  and
the residue functions (\ref{Ztdef}) and (\ref{Zldef}).
For completeness, we also give in each case the solution of (\ref{ne})
for the chemical potential $\mu$
as a function of the net electron density $n_e$.

\noindent {\bf Classical limit.}
In the classical limit, the plasma is nonrelativistic ($T << m_e$)
and nondegenerate ($m_e - \mu >> T$).
The Fermi distribution for electrons
can be approximated by the Boltzmann distribution $ e^{(\mu - E)/T}$,
and contributions from positrons can be ignored.
The net electron density, including the first correction
proportional to $T/m_e$, is
\begin{equation} {
n_e(T,\mu) \;=\;
e^{(\mu - m_e)/T} {1 \over \sqrt{2 \pi^3}} \left( m_e T \right)^{3/2}
\; \left( 1 \;+\; {15 \over 8} {T \over m_e} \right) \;.
} \label{nec} \end{equation}
This can be solved trivially for the chemical potential $\mu(T,n_e)$
as a function of the temperature and the net electron density.
The plasma frequency $\omega_p$ is
\begin{equation} {
\omega_p^2 \;=\;
e^{(\mu - m_e)/T} \sqrt{{8 \over \pi}} \; \alpha \left( m_e T^3 \right)^{1/2}
\; \left( 1 \;-\; {5 \over 8} {T \over m_e} \right) \;.
} \label{wplc} \end{equation}
Using (\ref{nec}) to eliminate $\mu$ in favor of the net electron density,
the expression (\ref{wplc}) reduces to
\begin{equation} {
\omega_p^2 \;=\; {4 \pi \alpha n_e \over m_e}
\left( 1 \;-\; {5 \over 2} {T \over m_e} \right) \;.
} \label{wplcn} \end{equation}
The polarization functions, including the first correction
proportional to $T/m_e$, are
\begin{equation} {
\Pi_t(\omega,k) \;=\; \omega_p^2
\left( 1 \;+\; {k^2 \over \omega^2} {T \over m_e} \right) \;,
} \label{Pitc} \end{equation}
\begin{equation} {
\Pi_\ell(\omega,k) \;=\; \omega_p^2
\left( {k^2 \over \omega^2} \;+\; 3 {k^4 \over \omega^4} {T \over m_e}
\right) \;.
} \label{Pilc} \end{equation}
The resulting dispersion equations are
\begin{equation} {
\omega_t^2 \;=\; k^2 \;+\; \omega_p^2
\left( 1 \;+\; {k^2 \over \omega_t^2} {T \over m_e} \right) \;,
\quad 0 \leq k < \infty \;,
} \label{wtc} \end{equation}
\begin{equation} {
\omega_\ell^2 \;=\; \omega_p^2
\left( 1 \;+\; 3 {k^2 \over \omega_\ell^2} {T \over m_e} \right) \;,
\quad 0 \leq k < \sqrt{1 + 3T/m_e} \; \omega_p \;.
} \label{wlc} \end{equation}
These are identical to the dispersion equations
(\ref{wtbps}) and (\ref{wlbps}) with $\omega_1 = \sqrt{5T/m_e} \; \omega_p$.
The transverse mass for the photon is $m_t = \sqrt{1 + T/m_e} \; \omega_p$.
The residue factors are
\begin{equation} {
Z_t(k) \;=\;
1 \;-\; {\omega_p^2 k^2 \over \omega_t^4} {T \over m_e} \;,
} \label{Ztc} \end{equation}
\begin{equation} {
Z_\ell(k) \;=\; {\omega_\ell^2 \over \omega_p^2}
\left(1 \;-\;  6 {k^2 \over \omega_\ell^2} {T \over m_e} \right) \;.
} \label{Zlc} \end{equation}

{\bf Degenerate limit.}
The degenerate limit is the limit of low temperature $T << \mu - m_e$.
In the limit $T = 0$, The Fermi distribution for electrons
reduces to a step function:  $n_F(E) = 1$ for $E < \mu$ and
$n_F(E) = 0$ for $E > \mu$.
The net electron density is
\begin{equation} {
n_e(T=0,\mu) \;=\; {1 \over 3 \pi^2} \; \left( \mu^2 - m_e^2 \right)^{3/2} \;.
} \label{ned} \end{equation}
The chemical potential at $T = 0$ is called the Fermi energy:  $\mu = E_F$.
The solution to (\ref{ned}) for the Fermi momentum
$p_F \equiv \sqrt{E_F^2 - m_e^2}$  as a function of $n_e$ is
\begin{equation} {
p_F \;=\; \left( 3 \pi^2 n_e \right)^{1/3} \;.
} \label{pfd} \end{equation}
The plasma frequency $\omega_p$ reduces in the $T = 0$ limit to
\begin{equation} {
\omega_p^2 \;=\; {4 \alpha \over 3 \pi} p_F^2 v_F \;,
} \label{wpldn} \end{equation}
where the Fermi velocity is $v_F = p_F/E_F$.
The polarization functions are
\begin{equation} {
\Pi_t(\omega,k) \;=\; \omega_p^2 \; {3\omega^2 \over 2 v_F^2 k^2}
\left( 1 - {\omega^2 - v_F^2 k^2 \over \omega^2}
{\omega \over 2 v_F k} \log {\omega + v_F k \over \omega - v_F k} \right) \;,
} \label{Pitdeg} \end{equation}
\begin{equation} {
\Pi_\ell(\omega,k) \;=\; \omega_p^2 \; {3 \over v_F^2}
\left( {\omega \over 2 v_F k} \log {\omega + v_F k \over \omega - v_F k} - 1
\right) \;.
} \label{Pildeg} \end{equation}
The dispersion equations reduce to (\ref{wtdeg}) and (\ref{wldeg}).
The transverse photon mass is
\begin{equation} {
m_t \;=\; \left[ {3 \over 2 v_F^2}
\left( 1 - {1 - v_F^2 \over 2 v_F} \log {1 + v_F \over 1 - v_F} \right)
\right]^{1/2} \omega_p \;.
} \label{mtd} \end{equation}
The inverses of the residue factors are
\begin{equation} {
Z_t^{-1} \;=\; 1 \;-\; {3 \over 2} {\omega_p^2 \over v_F^2 k^2}
\left( {3 \over 2} - {3 \omega_t^2 - v_F^2 k^2 \over 2 \omega_t^2}
{\omega_t \over 2 v_F k} \log {\omega_t + v_F k \over \omega_t - v_F k}
\right) \;,
} \label{Ztdeg} \end{equation}
\begin{equation} {
Z_\ell^{-1} \;=\; {3 \over 2} {\omega_p^2 \over v_F^2 k^2}
\left( {\omega_\ell^2 \over \omega_\ell^2 - v_F^2 k^2}
- {\omega_\ell \over 2 v_F k} \log {\omega_\ell + v_F k \over \omega_\ell -
v_F k} \right) \;.
} \label{Zldeg} \end{equation}
The dispersion equations (\ref{wtdeg}) and (\ref{wldeg}) can be used to
eliminate the logarithms from (\ref{Ztdeg}) and (\ref{Zldeg}).  The
resulting algebraic expressions for $Z_t$ and $Z_l$  have the form
(\ref{Ztstar}) and (\ref{Zlstar}), with $v_* = v_F$.

\noindent {\bf Relativistic limit.}
The relativistic limit is the limit of either high temperature
$T >> m_e$ or high density $\mu >> m_e$.
In the limit $m_e = 0$, the net electron density is
\begin{equation} {
n_e(T,\mu) \;=\; {1 \over 3 \pi^2} \mu
\left( \mu^2 \;+\; \pi^2 T^2 \right) \;.
} \label{ner} \end{equation}
The solution for the chemical potential as a function of $n_e$ is
\begin{equation} {
\mu(T,n_e) \;=\;
\left( \sqrt{ \left( {1 \over 2} p_F^3 \right)^2
		+ \left( {\pi^2 \over 3} T^2 \right)^3 }
	\;+\; {1 \over 2} p_F^3 \right)^{1/3}
\;-\; \left( \sqrt{ \left( {1 \over 2} p_F^3 \right)^2
		+ \left( {\pi^2 \over 3} T^2 \right)^3 }
	\;-\; {1 \over 2} p_F^3 \right)^{1/3} \;.
} \label{mur} \end{equation}
where $p_F = (3 \pi^2 n_e)^{1/3}$.
The plasma frequency is
\begin{equation} {
\omega_p^2 \;=\; {4 \alpha \over 3 \pi}
\left( \mu^2 \;+\; {1 \over 3} \pi^2 T^2 \right) \;.
} \label{wplr} \end{equation}
The polarization functions are
\begin{equation} {
\Pi_t(\omega,k) \;=\; \omega_p^2 \; {3\omega^2 \over 2 k^2}
\left( 1 - {\omega^2 - k^2 \over \omega^2}
	{\omega \over 2 k} \log {\omega + k \over \omega - k} \right) \;,
} \label{Pitrel} \end{equation}
\begin{equation} {
\Pi_\ell(\omega,k) \;=\; 3 \; \omega_p^2 \;
\left( {\omega \over 2 k} \log {\omega + k \over \omega - k} - 1 \right) \;.
} \label{Pilrel} \end{equation}
The dispersion equations reduce to (\ref{wtrel}) and (\ref{wlrel}).
The maximum plasmon momentum is infinite and the transverse mass
is $m_t = \sqrt{3/2} \; \omega_p$.
The inverses of the residue factors are
\begin{equation} {
Z_t^{-1} \;=\; 1 \;-\; {3 \over 2} {\omega_p^2 \over k^2}
\left( {3 \over 2} - {3 \omega_t^2 - k^2 \over 2 \omega_t^2}
	{\omega_t \over 2 k} \log {\omega_t + k \over \omega_t - k} \right) \;,
} \label{Ztr} \end{equation}
\begin{equation} {
Z_\ell^{-1} \;=\; {3 \over 2} {\omega_p^2 \over k^2}
\left( {\omega_\ell^2 \over \omega_\ell^2 - k^2}
- {\omega_\ell \over 2 k} \log {\omega_\ell + k \over \omega_\ell - k}
\right) \;.
} \label{Zlr} \end{equation}
The dispersion equations (\ref{wtrel}) and (\ref{wlrel}) can be used to
eliminate the logarithms from (\ref{Ztr}) and (\ref{Zlr}).  The resulting
algebraic expressions for $Z_t$ and $Z_l$ have the form
(\ref{Ztstar}) and (\ref{Zlstar}) with $v_* = 1$.

\subsection{Analytic Approximation}

The dispersion relations (\ref{wtdef}) and (\ref{wldef})
are integral equations and solving them numerically can be
computationally time-intensive.  It is therefore desirable to have
analytic approximations to these equations.  To derive such equations,
we begin from the expressions (\ref{Pit}) and (\ref{Pil}) for the
polarization functions and integrate by parts:
\begin{equation} {
Pi_t(\omega,k) \;=\; - {4 \alpha \over \pi} \int_0^\infty dp \; {p^3 \over E}
\left\{ {1 \over 2 v^2}
\left(  {\omega^2 \over k^2} - {\omega^2 - v^2 k^2 \over k^2}
{\omega \over 2 v k} \log {\omega + v k \over \omega - v k} \right)  \right\}
{d \ \over dp} \Bigg( n_F(E) \;+\; {\bar n}_F(E) \Bigg) \;,
} \label{Pitibp}	\end{equation}
\begin{equation} {
Pi_\ell(\omega,k) \;=\; - {4 \alpha \over \pi} \int_0^\infty dp \; {p^3 \over
E}
\left\{ {1 \over v^2}
\left({\omega \over 2 v k} \log {\omega + v k \over \omega - v k} - 1
\right)  \right\}
{d \ \over dp} \Bigg( n_F(E) \;+\; {\bar n}_F(E) \Bigg) \;.
} \label{Pilibp} \end{equation}
If the integrals over the momentum $p$ of the electrons and positrons
were dominated by a single velocity $v_*$, then the factors in the
curly brackets could be evaluated at $v = v_*$ and pulled outside the integral.
This is in fact the case in the classical limit, the relativistic limit,
and the degenerate limit.  In the classical limit, all the electrons have
velocities $v$ near 0.  In the relativistic limit, electrons and
positrons  all have velocities $v_* = 1$.  In the degenerate limit,
the factor $dn_F/dp$ is sharply peaked at the Fermi velocity $v_F$.
After pulling the expressions in curly brackets out of the integrands in
(\ref{Pitibp}) and (\ref{Pilibp}), the remaining
integrals over $p$ are proportional to the square of plasma frequency.
This can seen by using integration by parts on (\ref{wpl}):
\begin{equation} {
\omega_p^2 \;=\; - {4 \alpha \over 3 \pi} \;
\int_0^\infty dp \; {p^3 \over E}
{d \ \over dp} \Bigg( n_F(E) \;+\; {\bar n}_F(E) \Bigg) \;.
} \label{wplibp} \end{equation}
The resulting analytic expressions for the polarization functions are
\begin{equation} {
Pi_t(\omega,k) \;=\; \omega_p^2 \; {3 \over 2 v_*^2}
\left(  {\omega^2 \over k^2} - {\omega^2 - v_*^2 k^2 \over k^2}
	{\omega \over 2 v_* k} \log {\omega + v_* k \over \omega - v_* k} \right) \;,
} \label{Pitstar}	\end{equation}
\begin{equation} {
Pi_\ell(\omega,k) \;=\; \omega_p^2 \; {3 \over v_*^2}
\left({\omega \over 2 v_* k} \log {\omega + v_* k \over \omega - v_* k} - 1
\right) \;.
} \label{Pilstar} \end{equation}
Inserting these expressions into (\ref{wtdef}) and (\ref{wldef}),
we obtain the dispersion equations (\ref{wtstar}) and (\ref{wlstar}).

For suitable choices of the velocity $v_*$, the expressions (\ref{Pitstar})
and (\ref{Pilstar}) will be accurate in the classical, degenerate,
and relativistic limits.  The parameter $v_*$ can also be chosen so that
they are correct at small $k$ for all temperatures and electron densities.
At small $k$, the general expressions (\ref{Pit}) and (\ref{Pil})
for the polarization functions reduce to
\begin{equation} {
Pi_t(\omega,k) \;\approx\; {4 \alpha \over \pi} \int_0^\infty dp \;
{p^2 \over E}
\left\{ \left( 1 - {v^2 \over 3} \right) \;+\;
\left( {v^2 \over 3} - {v^4 \over 5} \right) {k^2 \over \omega^2} \right\}
\Bigg( n_F(E) \;+\; {\bar n}_F(E) \Bigg) \;,
} \label{Pitk}	\end{equation}
\begin{equation} {
Pi_\ell(\omega,k) \;\approx\; {4 \alpha \over \pi} \int_0^\infty dp \;
{p^2 \over E}
\left\{ \left( 1 - {v^2 \over 3} \right) \; {k^2 \over \omega^2} \;+\;
	\left( v^2 - {3 v^4 \over 5} \right) \; {k^4 \over \omega^4} \right\}
\Bigg( n_F(E) \;+\; {\bar n}_F(E) \Bigg) \;.
} \label{Pilk} \end{equation}
Expanding out the analytic expression (\ref{Pitstar})
for the transverse polarization function
in powers of $k$, we find that it agrees with
(\ref{Pitk}) to order $k^2$ provided that we take
$v_* = \omega_1/\omega_p$, where $\omega_1$ is given in (\ref{wone}).
We find also that the analytic expression (\ref{Pilstar})  for
the longitudinal polarization function
agrees with (\ref{Pilk}) to order $k^4$ for the same value of
$v_*$.

We can also obtain compact analytic expressions for the
residue factors $Z_t$ and $Z_\ell$.  Inserting (\ref{Pitstar}) and
(\ref{Pilstar}) into the formulas (\ref{Ztdef}) and (\ref{Zldef}),
we obtain
\begin{equation} {
Z_t^{-1} \;=\; 1 \;-\; {3 \over 2} {\omega_p^2 \over v_*^2 k^2}
\left( {3 \over 2} - {3 \omega_t^2 - v_*^2 k^2 \over 2 \omega_t^2}
{\omega_t \over 2 v_* k} \log {\omega_t + v_* k \over \omega_t - v_* k}
\right) \;,
} \label{Ztstari} \end{equation}
\begin{equation} {
Z_\ell^{-1} \;=\; {3 \over 2} {\omega_p^2 \over v_*^2 k^2}
\left( {\omega_\ell^2 \over \omega_\ell^2 - v_*^2 k^2}
- {\omega_\ell \over 2 v_* k} \log {\omega_\ell + v_* k \over
\omega_\ell - v_* k} \right) \;.
} \label{Zlstari} \end{equation}
The dispersion equations (\ref{wtstar}) and (\ref{wlstar}) can be used to
eliminate the logarithms from (\ref{Ztstari}) and (\ref{Zlstari}),
resulting in the simple algebraic expressions (\ref{Ztstar}) and
(\ref{Zlstar}).

\section{Effective photon-neutrino interaction}

The decay of a photon or plasmon into neutrino pairs proceeds through
an effective photon-neutrino interaction.  This effective interaction
arises from the electromagnetic coupling of a photon to electrons
or positrons in the plasma, together with the weak interaction coupling
of the electron or positron to a neutrino pair.
It can be summarized by an effective
vertex $\Gamma^{\alpha \mu}$ for the interaction of the photon field $A_\mu$
with the neutrino current ${\bar \nu} \gamma_\alpha (1 - \gamma_5) \nu$:
\begin{eqnarray}
\Gamma^{\alpha \mu}(\omega,{\vec k}) &=&
{G_F \over \sqrt{2}} {1 \over \sqrt{4 \pi \alpha}}
\Bigg( C_V \; Pi_\ell(\omega,k) \;
	\left( 1, {\omega \over k} {\hat k} \right)^\alpha
	\left( 1, {\omega \over k} {\hat k} \right)^\mu
\nonumber \\
&+& g^{\alpha i}
\left[ C_V \; Pi_t(\omega,k)  \left( \delta^{ij} - {\hat k}^i {\hat k}^j
\right)
\;+\; C_A \; \Pi_A(\omega,k)
\left( i \epsilon^{i j m} {\hat k}^m \right) \right]
g^{j \mu} \Bigg) \;,
 \label{veff} \end{eqnarray}
where $K^\mu = (\omega,{\vec k})$ is the 4-momentum of the photon.
This effective vertex satisfies the identity
$\Gamma^{\alpha \mu} K_\mu = 0$,
which guarantees the gauge invariance of the interaction.
The functions $Pi_t(\omega,k)$ and $Pi_\ell(\omega,k)$ in (\ref{veff}) are the
transverse and longitudinal electromagnetic polarization functions
defined in (\ref{Pitdef}) and (\ref{Pildef}).
The axial polarization function calculated to leading order in $\alpha$ is
\begin{equation} {
\Pi_A(\omega,k) \;=\;
8 \pi \alpha {K^2 \over k} \int {d^3p \over (2 \pi)^3} \; {1 \over 2 E}
\Bigg( n_F(E) \;-\; {\bar n}_F(E) \Bigg)
{ P \cdot K \; \omega - K^2 E
	\over (P \cdot K)^2 - (K^2)^2/4} \;,
} \label{Piax} \end{equation}
where $K^2 = \omega^2 - k^2$ and $P \cdot K = E \omega - {\vec p} \cdot
{\vec k}$.
The coefficients $C_V$ and $C_A$ depend implicitly on the neutrino type.
The vector coefficients are
\begin{equation} {
C_V \;=\; 2 \sin^2 \theta_W \;+\; {1 \over 2}
\quad {\rm for} \quad \nu_e \;,
} \label{CVe} \end{equation}
\begin{equation} {
C_V  \;=\; 2 \sin^2 \theta_W \;-\; {1 \over 2}
\quad {\rm for} \quad \nu_\mu, \; \nu_\tau \;,
} \label{CVmu} \end{equation}
where $\theta_W$ is the weak mixing angle:  $\sin^2 \theta_W \simeq 0.226$.
The axial vector coefficients are
\begin{equation} {
C_A \;=\; {1 \over 2}
\quad {\rm for} \quad \nu_e \;,
} \label{CAe} \end{equation}
\begin{equation} {
C_A \;=\; - {1 \over 2}
\quad {\rm for} \quad \nu_\mu, \; \nu_\tau \;.
} \label{CAmu} \end{equation}
The axial polarization function (\ref{Piax})
agrees with the expression used in Ref. \cite{kim},
provided that the nonrelativistic dispersion relation (\ref{wtarw})
is used to set $K^2 = \omega_p^2$.

For frequencies $\omega$ greater than $2 m_e$,
the expression (\ref{Piax}) for the axial polarization function
has an imaginary part that arises from the production of
$e^+ e^-$ pairs. As discussed in Appendix A,
this unphysical behavior can be eliminated without any loss of accuracy
by dropping the term $(K^2)^2/4$ in the denominator of  (\ref{Piax}).
The resulting expression for $\Pi_A(w,k)$,
when evaluated on the photon dispersion relation $\omega = \omega_t(k)$,
is still correct to leading order in $\alpha$.
Having dropped the $(K^2)^2/4$ term in the denominator of (\ref{Piax}),
the axial polarization function reduces to
\begin{equation} {
\Pi_A(\omega,k) \;=\; {2 \alpha \over \pi} {\omega^2 - k^2 \over k}
\int_0^\infty dp \; {p^2 \over E^2}
\left( {\omega \over 2 v k} \log {\omega + v k \over \omega - v k}
	\;-\; {\omega^2 - k^2 \over \omega^2 - v^2 k^2} \right)
\Bigg( n_F(E) \;-\; {\bar n}_F(E) \Bigg) \;.
} \label{PiA} \end{equation}
Its behavior for small $k$ is $\Pi_A(\omega,k) \rightarrow \omega_A k$, where
$\omega_A$ is given in (\ref{wA}).

The expression (\ref{PiA}) for the axial polarization function
can be evaluated analytically in the classical limit, the degenerate
limit, and the relativistic limit:

\noindent{\bf Classical limit.}
At leading order in $T/m_e$, the axial polarization function reduces to
\begin{equation} {
\Pi_A(\omega,k) \;=\; e^{(\mu - m_e)/T} \; \sqrt{{2 \over \pi}} \; \alpha \;
\left( {T^3 \over m_e} \right)^{1/2} \; {k (\omega^2 - k^2) \over \omega^2}
\;.
} \label{PiAc} \end{equation}
Eliminating the chemical potential and using the expression (\ref{wplc})
for the plasma frequency in the classical limit, (\ref{PiAc})
simplifies further to
\begin{equation} {
\Pi_A(\omega,k) \;=\; {\omega_p^2 \over 2 m_e} \; {k (\omega^2 - k^2) \over
\omega^2} \;.
} \label{PiAcl} \end{equation}

\noindent {\bf Degenerate limit.}
At $T = 0$, the axial polarization function reduces to
\begin{equation} {
\Pi_A(\omega,k) \;=\; {2 \alpha \over \pi} \; p_F \; {\omega^2 - k^2 \over k}
\left( {\omega \over 2 v_F k} \log {\omega + v_F k \over \omega - v_F k}  - 1
\right) \;.
} \label{PiAd} \end{equation}
After setting $\omega = \omega_t(k)$, the logarithm can be eliminated
using the dispersion relation (\ref{wtdeg}).  The resulting expression
for $\Pi_A(\omega_t,k)$ is (\ref{PiAstar})
with $v_* = v_F$ and $\omega_A = (2 \alpha/3 \pi) p_F v_F^2$.

\noindent {\bf Relativistic limit.}
For $m_e = 0$, the axial polarization function reduces to
\begin{equation} {
\Pi_A(\omega,k) \;=\; {2 \alpha \over \pi} \; \mu \; {\omega^2 - k^2 \over k}
\left( {\omega \over 2 k} \log {\omega + k \over \omega - k}  - 1\right) \;.
} \label{PiAr} \end{equation}
After setting $\omega = \omega_t(k)$, the logarithm can be eliminated
using the dispersion relation (\ref{wtrel}).  The resulting expression
for $\Pi_A(\omega_t,k)$ is (\ref{PiAstar})
with $v_* = 1$ and $\omega_A = (2 \alpha/ 3 \pi) \mu$.

We can derive an analytic approximation to the axial polarization function
using the same methods that were applied to the transverse and longitudinal
polarization functions in Appendix A.   Starting with the expression
(\ref{PiA}) and integrating by parts, we obtain
\begin{equation} {
\Pi_A(\omega,k) \;=\; - {2 \alpha \over \pi} {\omega^2 - k^2 \over k}
\int_0^\infty dp \; {p^3 \over E^2}
\left\{ {1 \over v^2} \left( {\omega \over 2 v k} \log {\omega + v k \over
\omega - v k}
	\;-\; 1 \right) \right\}
{d \ \over dp} \Bigg( n_F(E) \;-\; {\bar n}_F(E) \Bigg) \;.
} \label{PiAibp} \end{equation}
In the classical, degenerate, and relativistic limits,
the integral is dominated by a single momentum $v_*$.
The factor in curly brackets can therefore be evaluated at $v_*$
and pulled outside the integral.  The remaining integral over $p$
is proportional to $\omega_A$, which after applying integration
by parts to the expression in (\ref{wA}) can be written in the form
\begin{equation} {
\omega_A \;=\; - {2 \alpha \over 3 \pi}
\int_0^\infty dp \; {p^3 \over E^2}
{d \ \over dp} \Bigg( n_F(E) \;-\; {\bar n}_F(E) \Bigg) \;.
} \label{wAibp} \end{equation}
Our final expression for the axial polarization function is
\begin{equation} {
\Pi_A(\omega,k) \;=\; \omega_A \; {\omega^2 - k^2 \over k} {3 \over v_*^2}
\left( {\omega \over 2 v_* k} \log {\omega + v_* k \over \omega - v_* k} \;-\;
 1 \right) \;.
} \label{PiAan} \end{equation}
This expression is not only correct for all $k$ in the classical, degenerate,
and relativistic limits, but it has the correct behavior
for small $k$ at all temperatures and electron densities.
The axial polarization function $\Pi_A(\omega_t,k)$ evaluated at the photon
dispersion relation can be further simplified by using
the dispersion relation (\ref{wlstar}) to eliminate
the logarithm in (\ref{PiAan}), resulting in the
analytic expression given in (\ref{PiAstar}).

\section{Decay rate of photon or plasmon}

In this Appendix, we calculate the rates $\Gamma_t(k)$ and $\Gamma_l(k)$
for a photon or plasmon of momentum $k$ to decay into neutrino pairs.
We begin by writing down the matrix element
${\cal M}$ for the decay into a neutrino and antineutrino of momenta
${\vec p}_1$ and ${\vec p}_2$:
\begin{equation} {
{\cal M} \;=\; {G_F \over \sqrt{2}}
\left( \Gamma^{\alpha \mu} \epsilon_\mu({\vec k},\lambda) \right)
\; {\bar u}({\vec p}_1) \gamma_\alpha (1 - \gamma_5) v({\vec p}_2) \;,
} \label{Mdef} \end{equation}
where $\epsilon_\mu({\vec k},\lambda)$ is the polarization 4-vector given in
(\ref{epst}) for photons or (\ref{epsl}) for plasmons,
$\Gamma^{\alpha \mu}$ is the effective photon-neutrino vertex
given in (\ref{veff}),
and ${\bar u}({\vec p}_1)$ and $v({\vec p}_2)$ are the spinors for the
neutrino and antineutrino.
The matrix element must be squared and integrated over the phase space of
the neutrino and antineutrino:
\begin{equation} {
\Gamma_\lambda(k) \;=\; {1 \over 2 \omega_\lambda(k)}
\int {d^3p_1 \over (2 \pi)^3} {1 \over 2 p_1}
\int {d^3p_2 \over (2 \pi)^3} {1 \over 2 p_2}
\; (2 \pi)^4 \delta^4(P_1 + P_2 - K) \; |{\cal M}|^2 \;.
} \label{Gdef} \end{equation}
The subscript $\lambda$ on $\Gamma$ and on $\omega$
is either $t$, corresponding to helicity
$\lambda = \pm 1$, or $\ell$, corresponding to helicity $\lambda = 0$.
In the delta function, $P_1 = (p_1,{\vec p}_1)$, $P_2 = (p_2,{\vec p}_2)$, and
$K = (\omega_\lambda(k),{\vec k})$ are the 4-momenta of the neutrino,
antineutrino,
and photon or plasmon, respectively.
In the expression (\ref{Gdef}),
we have assumed for simplicity that the number density of neutrinos
and antineutrinos is negligible.  Otherwise the phase space integrals
in (\ref{Gdef}) must be weighted by appropriate Pauli blocking factors.
The only dependence in the matrix element (\ref{Mdef})
on the momenta ${\vec p}_1$ and ${\vec p}_2$ is in the spinor factor.
After multiplying the spinor factor by its complex conjugate,
it can be expressed in the form of a Lorentz tensor:
\begin{eqnarray}
&{\bar u}({\vec p}_1) \gamma^\alpha (1 - \gamma_5) v({\vec p}_2)
{\bar v}({\vec p}_2) \gamma^\beta (1 - \gamma_5) u({\vec p}_1)&
\nonumber \\
& \;=\; 8
\left( P_1^\alpha P_2^\beta \;+\; P_2^\alpha P_1^\beta
	\;-\; P_1 \cdot P_2 g^{\alpha \beta}
	\;-\; i \epsilon^{\alpha \beta \mu \nu} P_{1 \mu} P_{2 \nu} \right)& \;.
 \label{spinor} \end{eqnarray}
In the absence of Pauli blocking factors, the integral
over the phase space of the neutrino and antineutrino
in (\ref{Gdef}) can be carried out analytically, with the result
\begin{eqnarray}
\int {d^3p_1 \over (2 \pi)^3} {1 \over 2 p_1}
\int {d^3p_2 \over (2 \pi)^3} {1 \over 2 p_2}
\; (2 \pi)^4 \delta^4(P_1 + P_2 - K)
&& {\bar u}({\vec p}_1) \gamma^\alpha (1 - \gamma_5) v({\vec p}_2)
\; {\bar v}({\vec p}_2) \gamma^\beta (1 - \gamma_5) u({\vec p}_1)
\nonumber \\
&& \;=\; {1 \over 3 \pi}
	\left(K^\alpha K^\beta - K^2 g^{\alpha \beta} \right) \;,
 \label{intp} \end{eqnarray}
where $K^2 = w_\lambda^2 - k^2$.  Since the effective vertex
$\Gamma^{\alpha \mu}$ satisfies the identity
$K_\alpha \Gamma^{\alpha \mu} = 0$, only the $K^2 g^{\alpha \beta}$ term in
(\ref{intp}) contributes to the decay rate (\ref{Gdef}).
The decay rate then reduces to
\begin{equation} {
\Gamma_\lambda(k) \;=\;
- \; {G_F^2 \over 12 \pi} \; {\omega_\lambda(k)^2 - k^2 \over
\omega_\lambda(k)}
\; \left( \Gamma^{\alpha \mu} \epsilon_\mu({\vec k},\lambda) \right)
\; \left( \Gamma_{\alpha \rho} \epsilon^\rho({\vec k},\lambda) \right)^* \;.
} \label{Gsimp} \end{equation}

To complete the calculation of the decay rate,
the effective vertex $\Gamma^{\alpha \mu}$ in (\ref{veff})
must be contracted with the appropriate polarization
4-vector $\epsilon^\mu({\vec k},\lambda)$
and the energy $\omega$ must be evaluated at the
corresponding dispersion relation $\omega_t(k)$ or $\omega_\ell(k)$.
For the plasmon, only the first term in (\ref{veff}) contributes.
By the plasmon dispersion equation (\ref{wldef}), $\Pi_l(\omega,k)$ can be
replaced by $k^2$ and the contraction of $\Gamma^{\alpha \mu}$
with the polarization vector (\ref{epsl}) reduces to
\begin{equation} {
\Gamma^{\alpha \mu}(\omega_\ell(k),{\vec k}) \; \epsilon_\mu({\vec k},0) \;=\;
C_V \; {G_F \over \sqrt{8 \pi \alpha}} \;
\sqrt{Z_\ell(k)} \; k \; \omega_\ell(k)
\left( 1, {\omega \over k} {\hat k} \right)^\alpha \,.
} \label{egl} \end{equation}
For the photon, it is the last term in the effective vertex
(\ref{veff}) that contributes.
By the photon dispersion equation (\ref{wtdef}), $\Pi_t(\omega,k)$ can be
replaced by $\omega_t(k)^2 - k^2$ and the contraction of $\Gamma^{\alpha \mu}$
with the polarization vector (\ref{epst}) reduces to
\begin{eqnarray}
\Gamma^{\alpha \mu}(\omega_t(k),{\vec k}) \; \epsilon_\mu({\vec k},\pm 1) &=&
{G_F \over \sqrt{8 \pi \alpha}} \sqrt{Z_t(k)}
\Bigg( C_V \; (\omega_t(k)^2 - k^2) \left( 0, {\vec \epsilon}_\pm({\vec k})
\right)^\alpha
\nonumber \\
&-& C_A \; \Pi_A(\omega_t(k),k)
\left( 0, i {\hat k} \times {\vec \epsilon}_\pm({\vec k}) \right)^\alpha
\Bigg) \;.
 \label{egt} \end{eqnarray}
Squaring the expression (\ref{egt}) and (\ref{egl}) and inserting into
(\ref{Gsimp}), the decay rates of the photon and plasmon into neutrino pairs
reduce to
\begin{equation} {
\Gamma_t(k) \;=\;
{G_F^2 \over 48 \pi^2 \alpha} \; Z_t(k) \; {\omega_t(k)^2 - k^2 \over
\omega_t(k)}
\Bigg( C_V^2 \; (\omega_t(k)^2 - k^2)^2 \;+\; C_A^2 \; \Pi_A(\omega_t(k),k)^2
\Bigg) \;.
} \label{Gph} \end{equation}
\begin{equation} {
\Gamma_\ell(k) \;=\;
C_V^2 \; {G_F^2 \over 48 \pi^2 \alpha}
\; Z_\ell(k) \; \omega_\ell(k) \left( \omega_\ell(k)^2 - k^2 \right)^2 \;.
} \label{Gpl} \end{equation}

The expressions (\ref{Gpl}) and (\ref{Gph}) for the decay rates
of photons and plasmons include only the effects of electrons and positrons
in the plasma.
It is straightforward to include also the effects of protons
provided that the plasma frequency remains small compared to 700 MeV,
so that form factor effects can be neglected.
The effects of protons on the dispersion relations and on the
effective neutrino-photon vertex must both be included.
The expressions (\ref{Gph}) and (\ref{Gpl}) for the decay rates
are replaced by
\begin{eqnarray}
\Gamma_t(k) \;=\;
{G_F^2 \over 48 \pi^2 \alpha} \; Z_t \; {\omega_t^2 - k^2 \over \omega_t}
\Bigg( & \left( C_V \; Pi_t^{(e)}(\omega_t,k)
	\;-\; h_V \; Pi_t^{(p)}(\omega_t,k) \right)^2 &
\nonumber \\
\;+\; & \left( C_A \; \Pi_A^{(e)}(\omega_t,k)
	\;-\; h_A \; \Pi_A^{(p)}(\omega_t,k) \right)^2 & \Bigg) \;'
\label{Gphep} \end{eqnarray}
\begin{equation} {
\Gamma_\ell(k) \;=\;
{G_F^2 \over 48 \pi^2 \alpha}
\; Z_\ell \; {\omega_\ell (\omega_\ell^2 - k^2)^2 \over k^4}
\left( C_V \; Pi_\ell^{(e)}(\omega_\ell,k)
	\;-\; h_V \; Pi_\ell^{(p)}(\omega_\ell,k) \right)^2  \;,
} \label{Gplep} \end{equation}
where $Pi_t^{(e)}$, $Pi_\ell^{(e)}$, and $\Pi_A^{(e)}$ are the contributions
to the transverse, longitudinal, and axial polarization functions
{}from electrons, while $Pi_t^{(p)}$, $Pi_\ell^{(p)}$, and $\Pi_A^{(p)}$ are
the corresponding contributions from protons.
The proton terms have the same form as the electron terms
(\ref{Pitstar}), (\ref{Pilstar}), and (\ref{PiAan}),
except that the parameters $\omega_p$, $v^*=\omega_1/\omega_p$, and $\omega_A$
are calculated using the integrals (\ref{wpl}), (\ref{wone}),
and (\ref{wA}) with the
electron mass $m_e$ replaced by the proton mass $m_p$.
The coefficients $h_V$ and $h_A$ that describe the interactions
of the proton with neutrinos are
\begin{equation} {
h_V \;=\; - 2 \sin^2 \theta_W \;+\; {1 \over 2}
\quad {\rm for} \quad \nu_e \;, \nu_\mu \;, \nu_\tau \;,
} \label{hV} \end{equation}
\begin{equation} {
h_V  \;=\; {1 \over 2} g_A
\quad {\rm for} \quad \nu_e \;, \nu_\mu, \; \nu_\tau \;,
} \label{hA} \end{equation}
where $g_A \approx 1.26$.
The dispersion equations (\ref{wtdef}) and (\ref{wldef}) must also be
modified to include the effects of protons:
\begin{equation} {
\omega_t^2 \;=\; k^2 \;+\; Pi_t^{(e)}(\omega_t,k) \;+\;
Pi_t^{(p)}(\omega_t,k) \;,
} \label{wtep} \end{equation}
\begin{equation} {
\omega_\ell^2 \;=\; {\omega_\ell^2 \over k^2}
\left( Pi_\ell^{(e)}(\omega_\ell,k) \;+\; Pi_\ell^{(p)}(\omega_\ell,k)
\right)\;.
} \label{wlep} \end{equation}
The residue factors $Z_t$ in (\ref{Gphep}) and $Z_\ell$ in (\ref{Gplep})
are then given by the expressions (\ref{Ztstari}) and (\ref{Zlstari}),
except that in addition to the electron term on the right side,
there is also a proton term of the same form but with appropriate values
for the parameters $\omega_p$ and $v_*$.

\vfill\eject

\vfill \eject

\noindent
{\Large {\bf Figure Captions}}
\begin{enumerate}
\item Dispersion relations $\omega(k)$ for photons (upper solid curve)
and plasmons (lower solid curve)
at temperature $T = 10^{11}$ K and proton mass density
$\rho/\mu_e = 10^{12} \; {\rm g/cm}^3$.
\item Transverse (T), longitudinal (L), and axial vector (A) components
of the neutrino emissivity (in units of ${\rm erg/s/cm}^3$) as a function
of the proton mass density $\rho/\mu_e$ (in units of ${\rm g/cm}^3$)
at the temperature $T = 10^{11}$ K (solid curves).
Also shown are the corresponding emissivities
calculated with the 0-temperature dispersion relations (dashed curves).
\end{enumerate}
\vfill \eject


\begin{thebibliography}{99}
%
\bibitem{arw}
{J.B. Adams, M.A. Ruderman, and C.-H. Woo,
	{\it Phys. Rev.} {\bf 129}, 1383 (1963);
C.L. Inman and M.A. Ruderman,
	{\it Phys. Rev.} {\bf 140}, 1025 (1965).}
%
\bibitem{eba}
{E. Braaten, {\it Phys. Rev. Lett.} {\bf 66}, 1655 (1991).}
%
\bibitem{tsy}
{V.N. Tsytovich, {\it Sov. Phys. JETP} {\bf 13}, 1249 (1961).}
%
\bibitem{bps}
{G. Baudet, V. Petrosian, and E.E. Salpeter,
	{\it Astrophys. J.} {\bf 150}, 979 (1967).}
%
\bibitem{dad}
{D.A. Dicus, {\it Phys. Rev.} {\bf D6}, 941 (1972).}
%
\bibitem{mki}
{H. Munakata, Y. Kohyama, and N. Itoh,
	{\it Astrophys. J.} {\bf 296}, 197 (1986);
N. Itoh, T. Adachi, M. Nakagawa, Y. Kohyama, and H. Munakata,
	{\it Astrophys. J.} {\bf 339}, 354 (1989);
	(erratum in {\it Astrophys. J.} {\bf 360}, 741 (1990)).}
%
\bibitem{sswmt}
{P.J. Schinder, D.N. Schramm, P.J. Wiita, S.H. Margolis, and D.L. Tubbs,
	{\it Astrophys. J.} {\bf 313}, 531 (1987).}
%
\bibitem{sil}
{V.P. Silin, {\it Sov. Phys. JETP} {\bf 11}, 1136 (1960).}
%
\bibitem{klw}
{V.V. Klimov, {\it Sov. Phys. JETP} {\bf 55}, 199 (1982);
	H.A. Weldon, {\it Phys. Rev.} {\bf D26}, 1394 (1982).}
%
\bibitem{imhk}
{N. Itoh, H. Mutoh, A. Hikjita, and Y. Kohyama,
	{\it Astrophys. J.} {\bf 395}, 622 (1992).}
%
\bibitem{janc}
{B. Jancovici, {\it Nuovo Cim.} {\bf 25}, 428 (1962).}
%
\bibitem{alt}
{T. Altherr, E. Petitgirard, and T. del Rio Gaztellurrutia,
	Annecy preprint ENSLAPP-A-412/92 (December 1992).}
%
\bibitem{ebb}
{E. Braaten,
	{\it Astrophys. J.} {\bf 392}, 70 (1992).}
%
\bibitem{tz}
{V.N. Tsytovich, {\it Sov. Phys. JETP} {\bf 18}, 816 (1964);
M.H. Zaidi, {\it Nuovo Cim.} {\bf 40}, 502 (1965).}
%
\bibitem{kim}
{Y. Kohyama, N. Itoh, and H. Munakata,
	{\it Astrophys. J.} {\bf 310}, 815 (1986).}
%
\bibitem{kkr}
{R. Kobes, G. Kunstatter, and A. Rebhan,
	{\it Phys. Rev. Lett.} {\bf 64}, 2992 (1990).}
%
\bibitem{gb}
{G. Barton, {\it Ann. Phys. (N.Y.)} {\bf 200}, 271 (1990);
	E. Braaten, Northwestern preprint NUHEP-TH-92-22 (October 1992).}
%
\end{thebibliography}
\end{document}